\documentclass{acm_proc_article-sp}
\usepackage{enumitem}
\usepackage{afterpage}
\usepackage[usenames, dvipsnames]{color}
\usepackage{listings}
\usepackage{xcolor}
\usepackage{graphicx}
\usepackage{epstopdf}
\usepackage{caption}
\usepackage{subcaption}
\usepackage{float}
\usepackage{placeins}
\usepackage{url}

\begin{document}

\title{5G mmWave Module for the ns-3 Network Simulator
\numberofauthors{1} 
%
\author{
%
%
\alignauthor
M. Mezzavilla, S. Dutta, M. Zhang, M. R. Akdeniz, S. Rangan\\
       \affaddr{NYU Polytechnic School of Engineering}\\
       \affaddr{2 MetroTech Center, 11211, Brooklyn, New York}\\
       \email{\{mezzavilla,sdutta,menglei,akdeniz,srangan\}@nyu.edu}
}
}
\maketitle

\begin{abstract}
The increasing demand of data, along with the spectrum scarcity, are motivating a urgent shift towards exploiting new bands. This is the main reason behind identifying mmWaves as the key disruptive enabling technology for 5G cellular networks. Indeed, utilizing new bands means facing new challenges; in this context, they are mainly related to the radio propagation, which is shorter in range and more sensitive to obstacles. The resulting key aspects that need to be taken into account when designing mmWave cellular systems are \emph{directionality} and \emph{link intermittency}. The lack of network level results motivated this work, which aims at providing the first of a kind open source mmWave framework, based on the network simulator ns-3. The main focus of this work is the modeling of customizable channel, physical (PHY) and medium access control (MAC) layers for mmWave systems. The overall design and architecture of the model are discussed in details. Finally, the validity of our proposed framework is corroborated through the simulation of a simple scenario.
\end{abstract}

\category{I.6.5}{Simulation and Modeling}{Model Development}[Modeling methodolgies]
\category{I.6.7}{Simulation and Modeling}{Simulation Support Systems}[Environments]

\terms{Simulation, Modeling, Architecture, Design, Performance, Cellular, Wireless}

\keywords{mmWave, 5G, Cellular, Channel, Propagation, PHY, MAC.} 

\section{Introduction}
The ever increasing demand of wireless cellular data has motivated researchers to investigate the potentials of millimeter wave communication for the 5th generation of cellular technology. A substantial body of literature is currently available discussing physical measurements and formulations for millimeter wave channels \cite{sundeep}, \cite{mustafa}. As a logical next step, we aim at studying how the upper layers of the communication network stack work over millimeter wave physical channels. In this work we aim to develop the first millimeter wave module for the ns-3 network simulator \cite{ns3} that can be used to quantitatively analyze the performance of transport and application layer protocols over millimeter wave last-mile links.

The ns-3 network simulator currently implements a wide range of network protocols across various layers of the communication network. Due to this it is a valuable tool for researchers working on cross-layer design. The ns-3 simulator already hosts modules for the simulation of WiFi, WiMAX and 3GPP-LTE networks. In this paper we propose the first ns-3 module for the simulation of millimeter wave based communication systems.

The simulation module described in this paper is designed to be a fully customizable model where the user can plug in various parameters, like carrier frequency, bandwidth, frame structure, etc., describing the behavior of the millimeter wave channel and devices. In fact, the aim of this work is to enable researchers to flexibly use this module for various scenarios without the need of altering the source code.

 The rest of the article is organized as follows. In Section \ref{secframeW}, we discuss the architecture of the mmW module. Section \ref{secPhy} discusses the modeling of the physical layer. Section \ref{maclayer} follows with a discussion on the medium access control (MAC) layer for our module. The interfacing of the various modules is discussed in Section \ref{SecSAP}. In Section \ref{simuRes}, we show  the results obtained for a simple simulation scenario. Finally, we list our future work items and conclude the paper in Section \ref{concl}.

\section{mmWave framework}
\label{secframeW}
Our framework includes a basic implementation of mmWave devices, which comprises the propagation and channel model, the physical (PHY) layer, and the MAC layer. The module completely is developed in C++.
 The design of this module is inspired by the ns-3 Lena module which, in our opinion, has a very robust architecture.
{\bf Fig. \ref{classdiagram} } gives the UML diagram of the most important classes in our mmWave module. We will provide details about the various classes and their interoperation in the following sections.

\afterpage{
\begin{figure*}[]
\includegraphics[width=\textwidth] {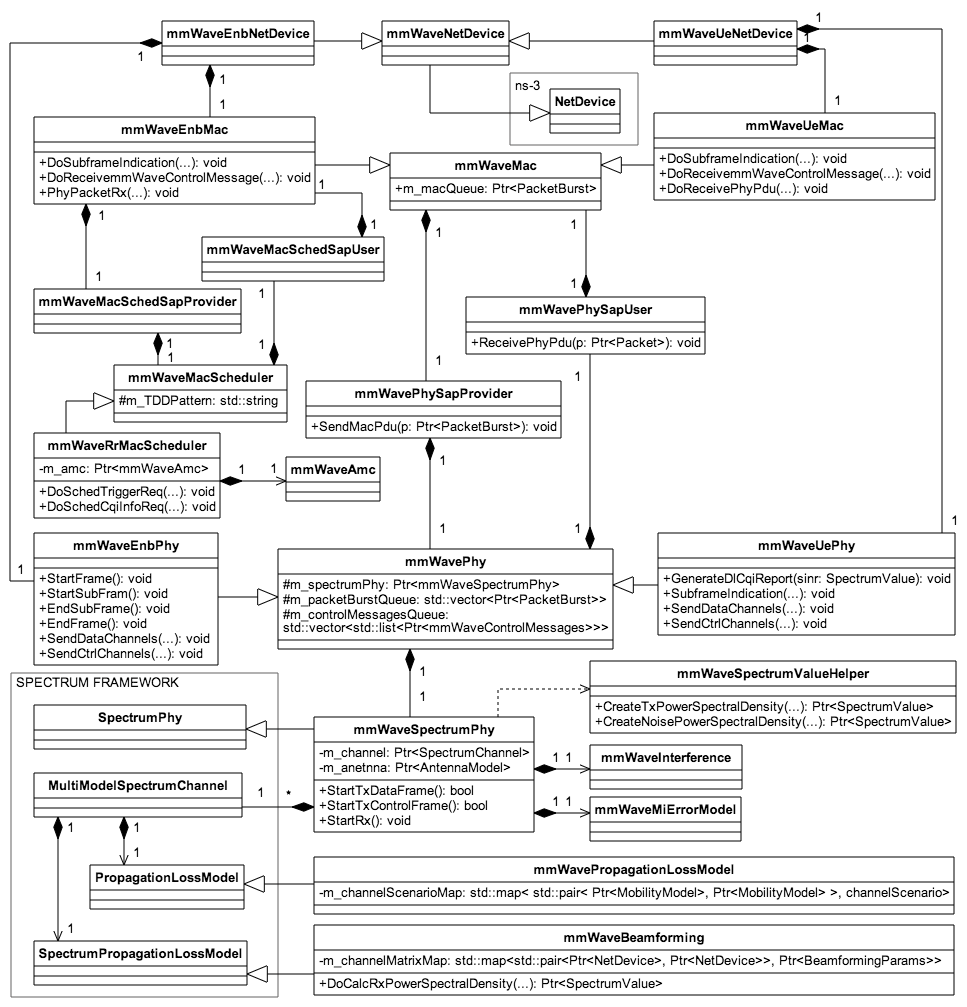}
\caption{\bf Class diagram for the mmWave module.}
\label{classdiagram}
\end{figure*}
\clearpage
}

\newpage

\section{PHY Layer}
\label{secPhy}
The salient features of the mmWave PHY layer are: (i) a fully customizable time division duplex (TDD) frame structure, (ii) a radio characterization that includes small and large scale channel variations, along with supporting multiple input multiple output (MIMO) techniques such as beamforming, (iii) a decoding error model at the receiver, (iv) an interference model, and (v) a feedback loop for channel adaptation. The following parts give a detailed outline of the implementation specifics of the PHY layer for the mmWave module.

\subsection{Frame Structure}
\label{subsec3.1}
The authors in \cite{ghosh2014millimeter} and \cite{radio_interface} contend that in order to reduce the latency over the air interface, the 5G mmWave systems will be targeted towards TDD operation. The ns-3 module for mmWave implements a customizable TDD frame structure. Each frame is subdivided into a number of subframes of fixed length specified by the user. Each subframe in turn is split into a number of slots of a fixed duration. Each slot comprises a specified number of OFDM symbols. A slot can be either {\bf control} or {\bf data}, assigned for either {\bf uplink} (UL) or {\bf downlink} (DL).

\begin{figure}[!t]
\includegraphics [scale=0.25, trim = 11mm 10mm 1mm 1mm,clip] {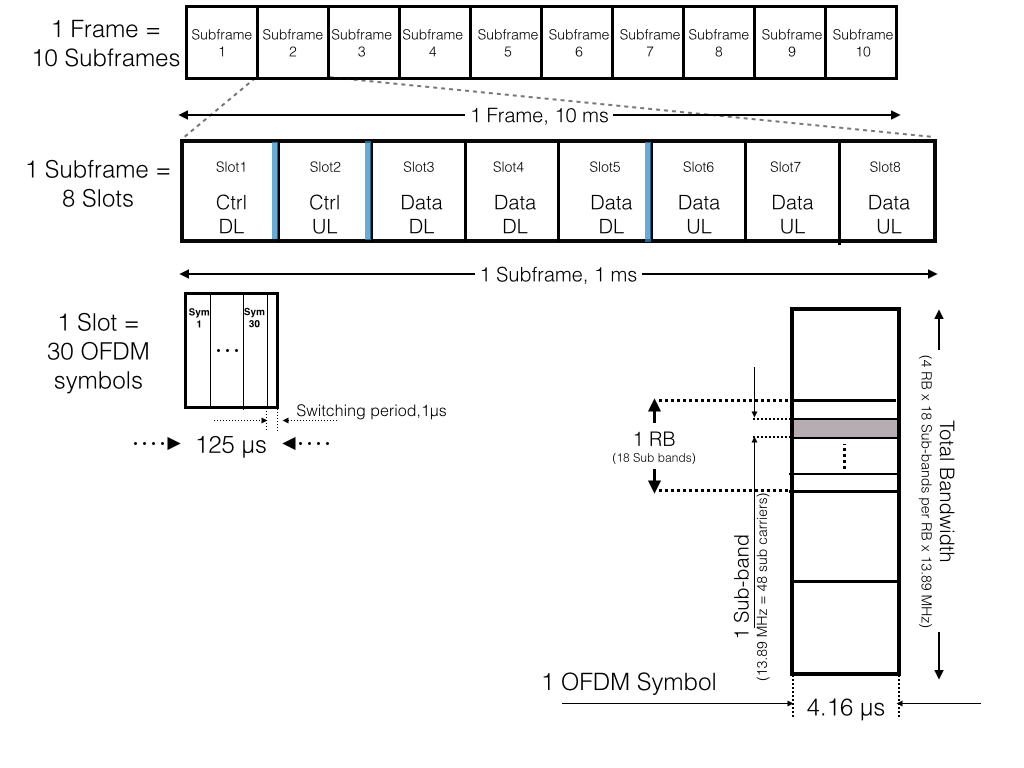}
\caption{\bf Example of mmWave frame structure.}
\label{fig1}
\end{figure}

{\bf Fig. \ref{fig1}} shows an example of the TDD frame structure based on the work in \cite{samsung}. Each frame of length 10ms is split in time into 10 subframes each of duration 1ms. Each subframe is further divided into 8 slots where each slot is of length $125 \mu s$ representing 30 orthogonal frequency division multiplexing (OFDM) symbols of length approximately $4.16 \mu s$. The first two slots are assigned for control in the downlink and in the uplink direction respectively. Slots 3 to 5 are allocated for downlink and slots 5 to 8 for uplink data transmission. A switching gap of $1 \mu s$ is introduced each time the allocated direction changes from uplink to downlink or vice versa. In the frequency domain the entire bandwidth of 1GHz is divided into 4 resource blocks (RBs). Each RB is subdivided into 18 sub-bands each of width 13.89 MHz making a total of 72 sub-bands for the entire bandwidth. Each of these sub-bands is composed of 48 sub-carriers. In this case, the total number of resource elements for one slot would be: $N_{RE} = 30\times (48\times 72) = 103680$.

\subsubsection{Parameter configuration}

The frame structure is completely customizable by the user. A common object of the \emph{mmwavePhyMacCommon} class stores the user specified values of all the parameters used by the simulator. The parameters used to customize the frame structure are specified in Table {\ref{table1}}.

The value of the Transmission Time Interval (TTI), which in our model is the duration of one slot, can be derived from the parameters in Table {\ref{table1}} as,
\begin{equation*}
\text{TTI} = SymbolPerSlot \times SymbolLength.
\end{equation*}
This is implemented by the function \emph{mmwavePhyMacCommon::GetTTI()}.

Similarly we can compute the bandwidth of one RB as,
\begin{equation*}
B_{RB} = SubbandsPerRB \times SubbandWidth.
\end{equation*}

\vspace{0.9mm}
This is implemented by the function \emph{mmwavePhyMacCommon::GetRBWidth()}.
The total system bandwidth can thus be computed as,
\begin{equation*}
B_{system} = B_{RB} \times NumResourceBlock.
\end{equation*}
This is implemented by the function \emph{mmwavePhyMacCommon::GetSystemBandwidth()}.

\begin{table*}[ht!]
\centering
\begin{tabular}{|| l | l | l ||}
\hline \hline
{\bf Parameter Name} & {\bf  Default Value} & {\bf Description} \\
\hline
{\it SymbolPerSlot} & 30 & Number of OFDM symbols per slot \\ \hline
{\it SymbolLength} & $4.16 \mu s$ & Length of one OFDM symbol in $\mu s$ \\ \hline
{\it SlotsPerSubframe} & 8 & Number of slots in one subframe \\ \hline
{\it SubframePerFrame }&  10 & Number of subframes in one frame \\ \hline
{\it NumReferenceSymbols} & 6 & The number of reference OFDM symbols per slot\\ \hline
{\it TDDControlDataPattern} & ``ccdddddd'' & The control (c) and data(d) pattern \\ \hline
{\it SubcarriersPerSubband} & 48 & Number of subcarriers in each sub-band \\ \hline
{\it SubbandsPerRB} & 18 & Number of sub-bands in one resource block \\ \hline
{\it SubbandWidth} & 13.89e6 & The width of one sub-band in  $Hz$ \\ \hline
{\it NumResourceBlock }&  4 & Number of resource blocks in one slot \\ \hline
{\it CenterFreq} & $28e9$ & The carrier frequency in $Hz$ \\
\hline \hline
\end{tabular}
\caption{\bf Parameters for configuring the mmWave frame structure.}
\label{table1}
\end{table*}

\subsubsection{Transmission schemes} 
The \emph{mmWaveEnbPhy} and the \emph{mmWaveUePhy} models the physical layer for the base station and the user device respectively. Broadly the physical layer (i) handles the transmission and reception of signals, (ii) simulate the start and the end of frames, subframes, and slots, (iii) deliver data packets and control messages received over the channel to the MAC layer, (iv) model the decoding error for the received signal and calculate the metrics like the signal to interference and noise ratio (SINR).

The physical layer contains one instance of the \emph{mmWaveSpectrumPhy} class. The transmission procedure for data frames is performed by the \emph{StartTxDataFrames}. For control messages the \emph{StartTxControlFrames} is invoked. The reception procedure is performed by \emph{StartRx} method. The functionality of the \emph{StartRx} is further subdivided into \emph{StartRxData} and \emph{StartRxControl}. Based on the total band of frequency available for transmission, the PHY computes the transmission power spectral density using the \emph{mmWaveSpectrumValueHelper} component and uses this value for the signal transmission. 

After the reception of the data packets, the PHY layer calculates the SINR of the received signal taking into account the MIMO beamforming gains. The physical layer at the user device maps the calculated SINR into a Channel Quality Indicator (CQI), which is fed-back to the base station for the resource allocation. Control signals are assumed to be ideally transmitted. As discussed in Section \ref{errormodel}, the PHY layer also incorporates the error model where a probabilistic approach is used to determine whether a packet should be dropped by the receiver. The correctly received packets are sent to the MAC layer using the service access point (SAP) discussed in Section \ref{SecSAP}.

The physical layer handles the start and the end of frames, subframes and slots based on the TTI derived from the user specified configurations. Based on the resource allocation scheme decided by the \emph{mmWaveEnbMac}, as described in Section \ref{SecMac}, the PHY decides the nature of the communication for a particular slot (data/control) and the direction of the message to transfer (uplink/downlink). At the beginning of each slot the eNodeB PHY sends a \emph{SubframeIndication} to the MAC. The subframe indication for the first slot triggers the scheduling and resource allocation functions. The data packets and the control messages received from the MAC are stored in the \emph{PacketBurstQueue} and the \emph{ControlMessageQueue}, respectively, and are transmitted to the connected device in the allocated slots.

\subsection{Channel Modeling}
\begin{figure} [b!]
\includegraphics [scale = 0.38]{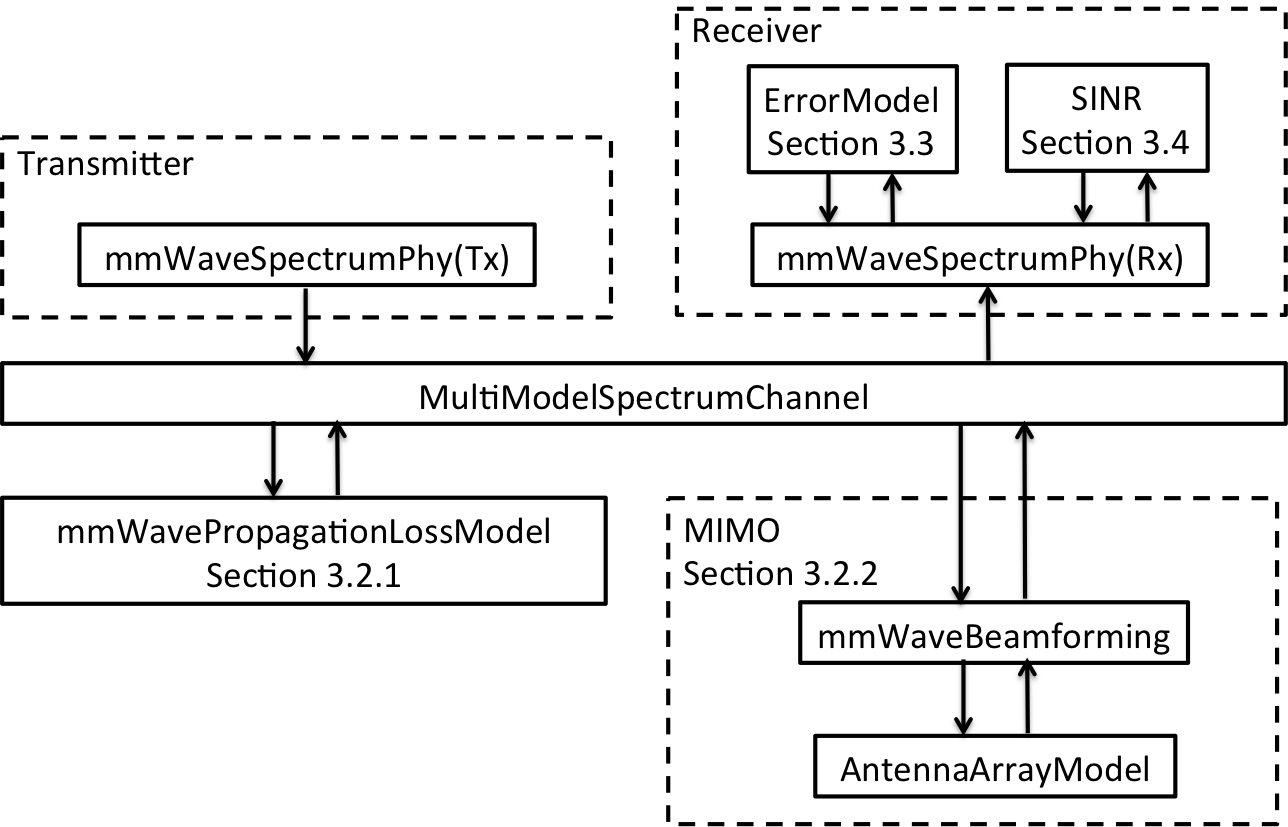}
\caption{\bf mmWave channel model.}
 \label{ChannelModel}
\end{figure}
As illustrated in Fig. \ref{ChannelModel}, we need to take into account a number of procedures to capture characteristics of the mmWave propagation. The key contribution here relates to the computation of the multi-antenna gains, which is particularly critical for mmWave communications. 

The link budget  for the mmWave propagation channel is given by,
\begin{equation}
P_{RX}=P_{TX}+G_{BF}-PL-SW,
\end{equation}
where $P_{RX}$ is the total received power in dBm,  $P_{TX}$ is the total transmit power, $G_{BF}$ is the beamforming gain, and finally $PL$ and $SW$ represent the pathloss and shadowing, respectively.

\subsubsection{mmWave Propagation Loss Model}

The mmWave pathloss model can be modeled with $3$ states, as reported in \cite{mustafa}, line of sight(LOS), non-line of sight (NLOS) and outage.
For each link, the channel is determined through the following procedure: 
\begin{itemize}
\item based on the distance between the UE and the eNB, determine the probability of the link being in each of the three states using the model in \cite{mustafa} $(P_{LoS}, P_{NLoS}, P_{out})$;
\item uniformly pick a reference value $(P_{REF})$ between 0 and 1 and compare with the probability associated with each channel state;
\item {\it if} $P_{REF} \leq  P_{LOS}$, pick LOS channel; 

{\it else if} $P_{LOS} < P_{REF} \leq  P_{LOS} + p_{NLOS}$, pick NLOS channel; 

{\it otherwise} pick outage.
\end{itemize}
For each link, on determining the channel state, the pathloss and shadowing is obtained by, 
\begin{equation}
PL(d)[dB] = \alpha+\beta10log_{10}(d)+\xi, \quad
\xi \sim N(0,\sigma^{2}),
\end{equation}
where $\xi$ represent shadowing, parameter $d$ represents the distance from receiver to transmitter, the values of parameter $\alpha$, $\beta$, and $\sigma$ for each channel scenario are given in \cite{mustafa}.

\subsubsection{MIMO}
Due to the high pathloss, multiple antenna with beam forming is essential to ensure acceptable range of communication in mmWave systems. We briefly discuss the associated concepts of channel matrices and beam forming in this respect.

{\bf Channel matrix:} We model the mmWave channel as a combination of clusters, each composed of several subpaths.\footnote{See \cite{mustafa} for a long term statistical characterization of the mmWave channel.} The channel matrix is described by the following:
\begin{equation}
H(t,f)=\sum_{k=1}^{K}\sum_{l=1}^{L_k}g_{kl}(t,f){\bf u}_{rx}(\theta^{rx}_{kl},\phi^{rx}_{kl}){\bf u}^*_{tx}(\theta^{tx}_{kl},\phi^{tx}_{kl})
\end{equation}
where, {$K$} is the number of clusters, {$L_k$} the number of subpaths in cluster k, {$g_{kl}(t,f)$} the small-scale fading over frequency and time; {$\bf{u}_{rx}(\cdot)$}is the spatial signature of the receiver, {$\bf{u}_{tx}(\cdot)$} the spatial signature of the transmitter.

The small-scale fading is generated based on the number of clusters, number of subpaths per cluster, Doppler shift, power spread, delay spread and angle of arrival (AoA) as given in \cite{mustafa} by,
\begin{equation}
g_{kl}(t,f)=\sqrt{P_{lk}}e^{2\pi if_{d}cos(\omega_{kl})t-2\pi i\tau _{kl}f}
\label{eq:smallscale}
\end{equation}
where, {$P_{kl}$} is the power spread; {$f_{d}$} is the maximum Doppler shift; {$\omega_{kl}$} is the AoA of the subpath relative to the direction of motion; {$\tau _{kl}$} gives the delay spread, and {$f$} is the carrier frequency.

{\bf Beamforming:} In order to support phased-array antennas, a new \emph{AntennaArrayModel} class is developed, which contains a complex beamforming vector. For both transmitter and receiver, based on the channel matrices, the beamforming vectors are computed using the power algorithm.

The beamforming gain from transmitter $i$ to receiver $j$ is given as,
\begin{equation}
G(t,f)_{ij}= |{\bf w}^{*}_{rx_{ij}}{\bf H}(t,f)_{ij}{\bf w}_{tx_{ij}}|^{2}
\label{eq:channelgain}
\end{equation}
where, {${\bf H}(t,f)_{ij}$} is the channel matrix of $ij^{th}$ link, {${\bf w}_{tx_{ij}}$} is the beamforming vector of transmitter $i$, when transmitting to receiver $j$ and {${\bf w}_{rx_{ij}}$} the beamforming vector of receiver $j$, when receiving from transmitter $i$.

\subsubsection{Channel Configuration}
To reduce the computational complexity, the channel matrices and beamforming vectors are pre-generated in  $\text{\textsc MATLAB}^{\text{\textregistered}}$.

{\bf Load Files:}
At the beginning of each simulation we load 100 instances of the spatial signature matrices, along with the beamforming vectors. Then, a channel matrix instance per UE-eNodeB pair is randomly picked to characterize the radio link. As we will discuss later, we simulate the long term fading by randomly picking an instance of the channel form the pre-generated files.

{\bf Link Initialization and Channel Matrix Updates:}
In the \emph{mmWaveBeamForming} class we define the a member \emph{m\_channelMatrixMap} ({\bf Fig. \ref{classdiagram}}) to map the channel matrix instance to each radio link. During the simulation, the small-scale fading is calculated at every slot, based on Eq.\ref{eq:smallscale}. The speed of the user is obtained directly from the mobility model. The remaining parameters are only subject to the environmental conditions , rural, urban etc., and therefore assumed constant over the entire simulation time. The reference values for different environments were recorded during the our mmWave channel propagation measurement campaign reported in \cite{measurements}. On the other hand, for the large-scale fading, the spatial signature matrices are periodically updated with customizable interval, say 100 ms, to simulate a sudden change of the perceived channel.

\begin{figure} [t!]
\includegraphics [width=90mm, trim = 18mm 24mm 1mm 60mm, clip]{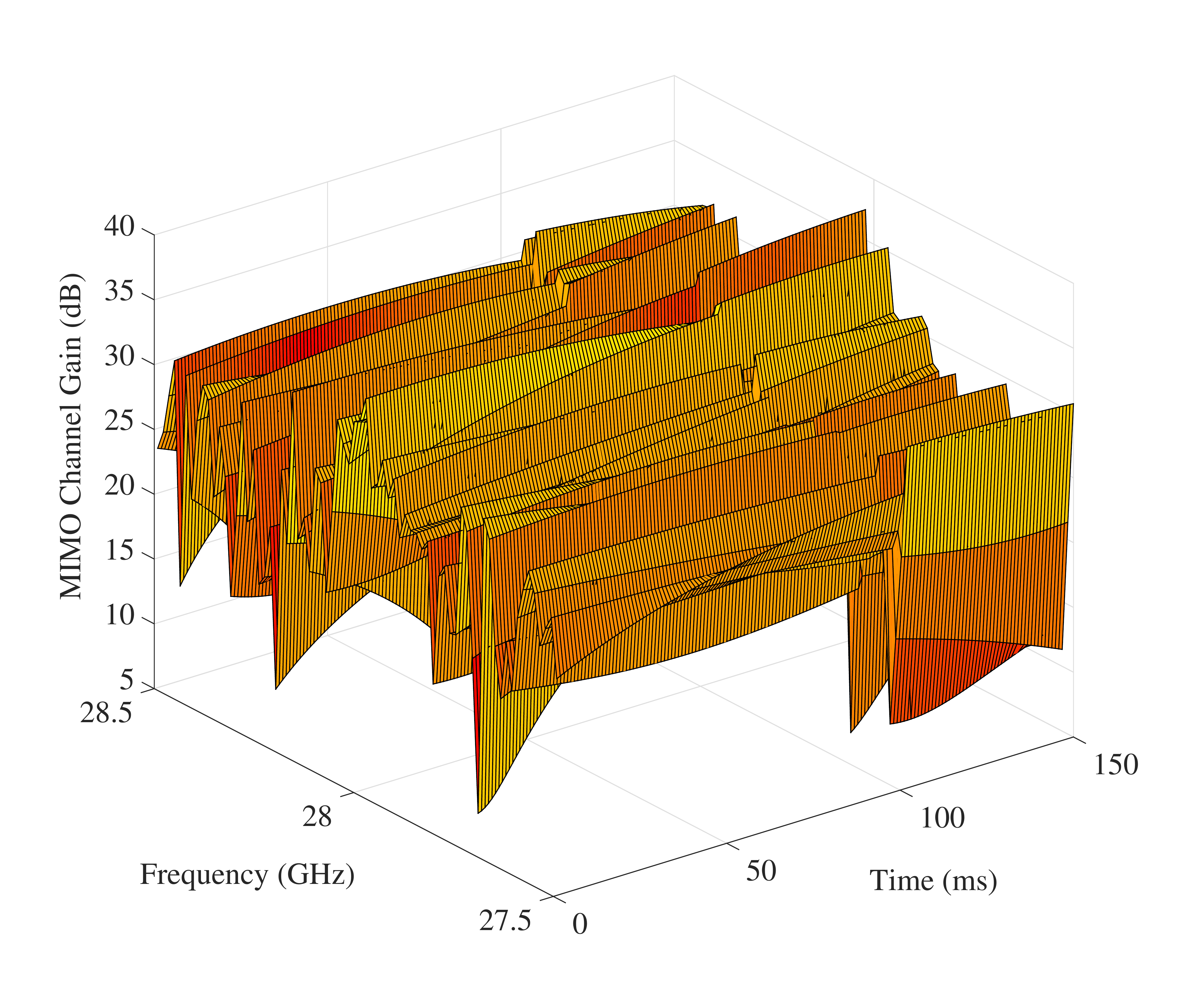}
\caption{\bf MIMO channel gain over frequency and time.}
\label{ChannelGain}
\end{figure}

{\bf Beamforming Vector:}
When configuring each radio link, the beamforming vector is stored in the antenna array model of both base station and user side; the former will store the beamforming vectors of all the  UEs, the latter will only store the beamforming vectors of the base stations within range. 

The beamforming parameters are defined using following structures:

\lstset {language=C++}
\begin{lstlisting}
struct BeamFormingParams
{
  complexVector_t  m_txW; 
  complexVector_t  m_rxW; 
  ChannelMatrix    m_channelMatrix;
}
struct ChannelMatrix
{
  complex2DVector_t  m_txSpatialMatrix;
  complex2DVector_t  m_rxSpatialMatrix;
}
\end{lstlisting}

We can observe in {\bf Fig. \ref{ChannelGain}} the channel gain trend obtained in a scenario where the number of antennas at the base station and mobile device is 64 and 16, respectively, and the user is moving at a speed of 36 km/h. The beamforming gain with small-scale fading is calculated using on Eq. \ref{eq:channelgain}. The small-scale fading includes two components, frequency selective fading and time selective fading, caused by multipath effect and Doppler effect, respectively. The spatial signatures and beamforming vectors are periodically updated with larger interval to capture the effects of long term fading which cause the sudden drop of beamforming gain around 100 ms.

\subsection{Error Model}
\label{errormodel}
Similar to ns-3 Lena module, our mmWave module includes a error model for data packets according to the standard link-to-system mapping (LSM) techniques. By utilizing the LSM and the Mutual Information Based Effective SINR (MIESM) \cite{marco}, the receiver computes the error probability for each transport block (TB) and determines whether the packet can be decoded or not. The TB can be composed of multiple codeblocks (CB) and its size depends on the channel capacity. The block error probability (BLER) of each CB depends on its size and associated MCS:
\begin{equation}
C_{BLER,i}(\gamma_i)=\frac{1}{2}\left [ 1-\textup{erf}\left ( \frac{\gamma_i -b_{C_{SIZE},MCS}}{\sqrt{2} c_{C_{SIZE},MCS}} \right ) \right ],
\end{equation}
where $\gamma_i$ is the mean mutual information per coded bit (MMIB) of the codeblock $i$, $b_{C_{SIZE},MCS}$ and $c_{C_{SIZE},MCS}$ corresponds to the mean and standard deviation of the Gaussian cumulative distribution, respectively. Now we can compute the TB block error rate: 
\begin{equation}
T_{BLER}=1-\prod_{i=1}^{C}(1-C_{BLER,i}(\gamma_i)).
\end{equation}
In case of failure, the PHY layer does not forward the incoming packet to the upper layers and, at the same time, triggers a retransmission process.\footnote{This can be a TCP retransmission, or an hybrid automatic repeat request (HARQ), which is part of our future work.}

\subsection{Interference}
\label{interference}
Albeit being presumably less threatening in the mmWave regime, because of the directionality of the multiantenna propagation, interference computation is still pretty relevant in terms of system level simulations. In fact, there might be some special spatial cases where interference is non negligible. Therefore, we propose an interference computation scheme that takes into account the beamforming directions associated with each link. 

\begin{figure} [t!]
\includegraphics [width=\columnwidth]{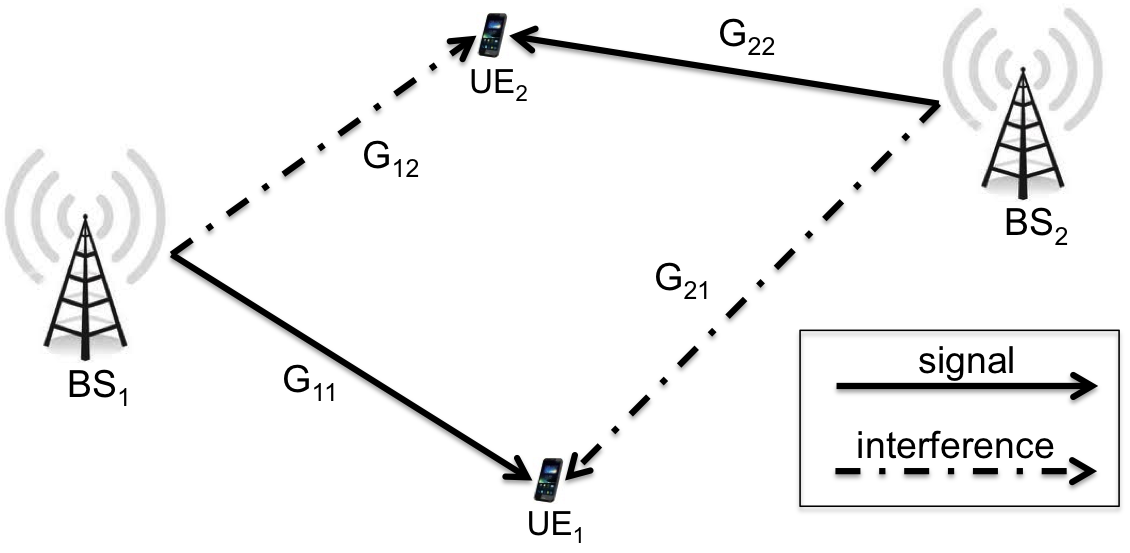}
\caption{\bf Interference model.}
\label{fig:interference}
\end{figure}

We will use {\bf Fig. \ref{fig:interference}} as a reference. As an example, we compute the SINR between $BS_1$ and $UE_1$. To do so, we first need to obtain the channel gains associated with both the desired and interfering signals. 
Following Eq. \ref{eq:channelgain}, we get
\begin{equation}
\begin{aligned}
G_{11}= |{\bf w}^{*}_{rx_{11}}H(t,f)_{11}{\bf w}_{tx_{11}}|^{2}, &
\\ G_{21}= |{\bf w}^{*}_{rx_{11}}H(t,f)_{21}{\bf w}_{tx_{22}}|^{2}.
\end{aligned}
\end{equation}

We can now compute the SINR:
\begin{equation}
SINR_{11}= \frac{\frac{P_{Tx,11}}{PL_{11}}G_{11}}{\frac{P_{Tx,22}}{PL_{21}}G_{21}+BW\times N_0},
\end{equation}
where $P_{Tx,11}$ is the transmit power of $BS_1$, $PL_{11}$ is the pathloss between between $BS_1$ and $UE_1$, and $BW\times N_0$ is the thermal noise. 

\subsection{CQI Feedbacks}
\label{cqifeedbacks}
In order to ensure reliable communication over a variable channel, feedback mechanisms are key in mostly all cellular communication systems. Similar to LTE we utilize the CQI feedback scheme for our module.
The downlink CQI feedback message is generated by the \emph{mmWaveUePhy}. In our simulator the UE computes the CQI based on the SINR of the signal received in a particular data slot. The computation of CQI is the same as that for LTE as given in \cite{FemtoForum} and \cite{ns3LTE}.

\section{MAC Layer}
\label{maclayer}
The MAC layer is developed using the class \emph{mmWaveMac} which is the base class for the \emph{mmWaveEnbMac} for the eNodeB and the \emph{mmWaveUeMac} for the user. The chief function of this layer is to deliver data packets coming from the upper layers (the net device in this case) to the physical layer and vice-versa. In fact this layer is designed for the synchronous delivery of upper layer data packets to the PHY layer which is key for proper data transfer in TDD mode.

The eNodeB MAC layer is connected to the scheduler module using the MAC-SCHED service access point (Sec. \ref{SecSAP2}). The relationship between the PHY, MAC and the scheduler module for the eNodeB is depicted in Fig. \ref{fig2}. Thus the MAC layer communicates the scheduling and the resource allocation decision to the PHY layer. The scheduler hosts the adaptive modulation and coding (AMC) module. 
The following sub-sections discuss these features in depth.

\subsection{Adaptive Modulation and Coding }
\label{SecAMC}
The working of the AMC is similar to that for LTE. The user measures the CQI for each downlink data slots it is allocated. The CQI information is then forwarded to the eNodeB using the \emph{mmWaveCqiReport} control message. The eNodeB scheduler uses this information to compute the most suitable modulation and coding scheme for the communication link.

The AMC is implemented by the eNodeB MAC schedulers. During resource allocation, for the current frame work, the wide band CQI is used to generate the modulation and coding scheme (MCS) to be used and the transport block (TB) size that can be transmitted over the physical layer. The AMC module provides this frame work for the unique mapping of the CQI, the MCS, spectral efficiency and the TB size. The TB size is calculated based on the values of the total number of subcarriers per resource block derived from the user customized configuration, the number of symbols per slot and the number of reference symbols per slot. A cyclic redundancy code (CRC) length of 24 bits is used.

\subsection {Scheduler}
\label{scheduler}
Following the design strategy for the ns-3 LTE module \cite{ns3LTE}, the virtual class \emph{mmWaveMacScheduler} defines the interface for the implementation of MAC scheduling techniques. The scheduler performs the scheduling and resource allocation for a subframe with both downlink and uplink slots.

\subsubsection{TDD scheme}
\label{tddscheme}
The TDD scheme enforced by the scheduler module is based on the user specified parameter \emph{``TDDControlDataPattern''} given in Table \ref{table1}. The slots specified for control are assigned alternately for downlink and uplink control channels. The data slots are equally divided between downlink and uplink slots with the first $n/2$ data slots allocated to downlink data and rest for uplink, where $n$ is the total number of data slots. This scheme minimizes the switching time between uplink and downlink data transmissions.

The scheme described in Fig. \ref{fig1} is an example of the implementation of the above algorithm with the default control data pattern. This module will be further enhanced in future to incorporate more advanced features like dynamic TDD.
 
\subsubsection{Resource Allocation}
Using the \emph{mmWavePhyMacCommon} object, the division of resources in the frequency domain can be customized as given in Table \ref{table1}.

The MAC scheduler currently implements a simple round robin algorithm to allocate uplink and downlink data slots to the connected users. All the frequency elements in a particular slot are assigned to the same user.  The control slots are not allocated to any particular user. Any or all user can receive from and transmit to the base station in the control slots. 

\begin{figure}[t!]
\includegraphics [scale=0.33, trim = 53mm 70mm 30mm 65mm,clip] {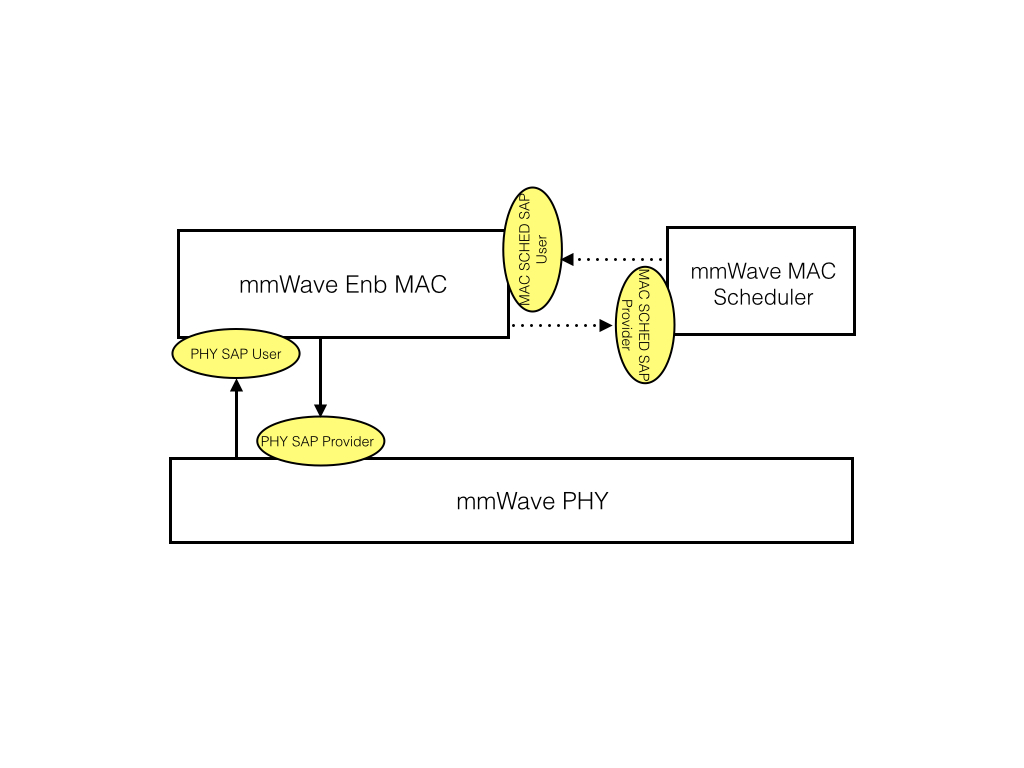}
\caption{\bf PHY, MAC and scheduler modules with the associated SAPs.}
\label{fig2}
\end{figure}
\label{SecMac}

For the case of the round robin scheduler the use of the CQI is limited to the determination of the transport block size. In future the CQI information will be used to actively control the scheduling decisions.

\section{Service access points}
\label{SecSAP}
The interface between the PHY and the MAC and the MAC and the scheduler are defined as service access points (SAPs) as given in \cite{FemtoForum}. The relationship between the PHY, MAC and the scheduler through their associated SAPs are shown in {\bf Fig. \ref{fig2}}. The relationship between modules connected through SAPs can be viewed as that of service providers and users. The SAP provider caters to the requirement of the SAP user based on certain requests received from the user. 

\subsection{PHY--MAC}
The communication between the MAC and the PHY layer using the MAC-PHY SAP is through the following processes:
\begin{itemize} [leftmargin=*]  \vspace{-3mm}
\item[i] {\bf Subframe Indication}: The subframe indication is sent by the PHY layer to the MAC at the beginning of each slot (unlike LTE where it is sent every subframe). The subframe indication for slot 1 for a particular subframe triggers the scheduling procedure for the eNodeB MAC. The subsequent indications are required for proper delivery of upper layer data.
\item[ii] {\bf Data transmission}: The eNodeB MAC maintains data queues for each of the connected UE and just one such queue is sufficient for the user device. Based on the scheduling scheme and the allocated resources, the MAC layer will send the scheduled number of packets (given by the transport block size) to the PHY layer for transmission over the radio link.
\item[iii] {\bf Scheduling and allocation notifications}: The scheduling and resource allocation decision received by the eNodeB MAC from the scheduler is relayed to the PHY layer using the \emph{mmWaveResourceAllocation} message. The PHY of the base station in turn transmits this message to all the connected users notifying all the attached devices of the scheduling decision. 
\item[iv] {\bf CQI notification}: Based on the SINR of the received data slots, the UE PHY calculate the CQI and transmits it to the base station in the next uplink control slot. The eNodeB PHY on receiving the \emph{mmWaveCqiReport} control message, relays it to the MAC.
\end{itemize}

\subsection{MAC--SCHED}
\label{SecSAP2}
The eNodeB MAC uses the service provided by the scheduler by the following processes:
\begin{itemize} [leftmargin=*]  \vspace{-3mm}
\item[i] {\bf Trigger Request and Configuration Indication}: On receiving the Subframe Indication for slot 1 of a particular subframe, the MAC sends a \emph{Scheduling Trigger Request} to the scheduler for the $(subframeNum+ delay)^{th}$ subframe, where the delay is specified by the user using the parameter \emph{L1L2ControlLatency}. The scheduler returns the scheduling and allocation decisions in the \emph{Scheduling Configuration Indication} in response to the trigger.
\item[ii] {\bf CQI notification}: The eNodeB MAC on receiving the CQI information from the PHY, sends it to the scheduler. The scheduler needs this information for future scheduling decisions.
\end{itemize}

\begin{figure*}[t!]
        \centering
         \begin{subfigure}[b]{0.4\textwidth}
                \includegraphics[width=\textwidth]{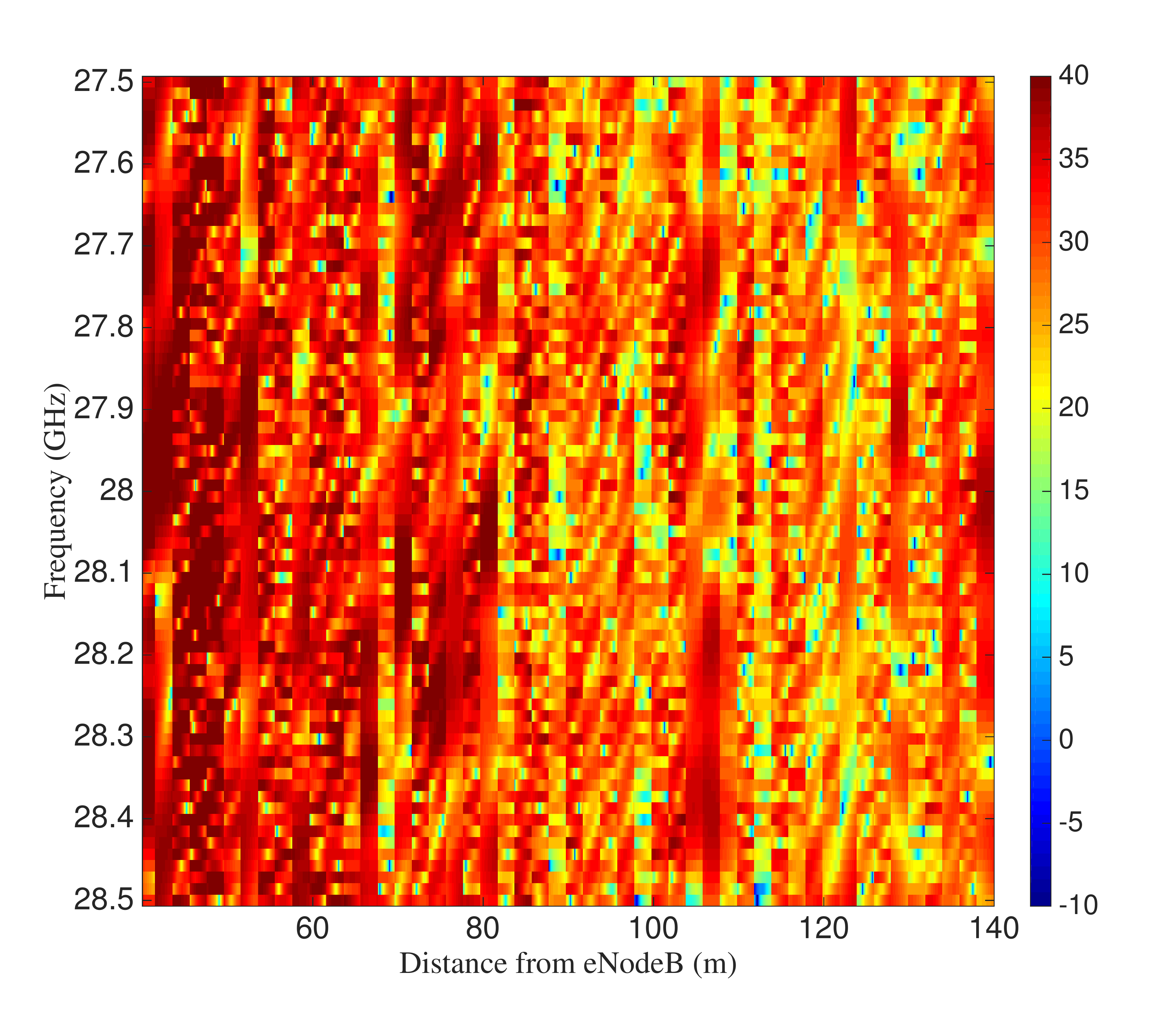}
                \caption{SINR for user with LoS link.}
                \label{SinrLos}
        \end{subfigure}
        \begin{subfigure}[b]{0.4\textwidth}
                \includegraphics[width=\textwidth]{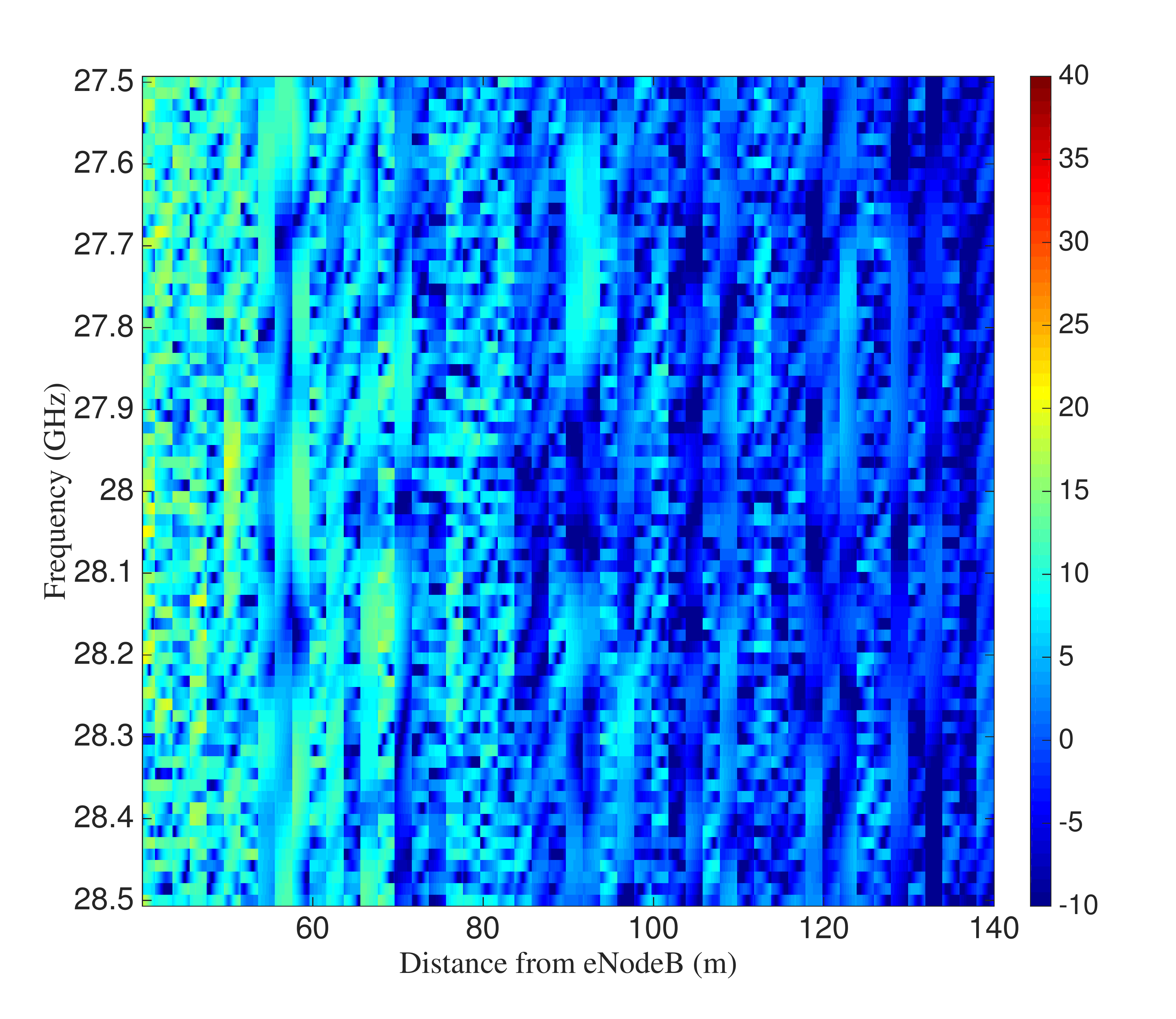}
                \caption{SINR for user with NLoS link.}
                \label{SinrNlos}
        \end{subfigure}
        \begin{subfigure}[b]{0.4\textwidth}
                \includegraphics[width=\textwidth]{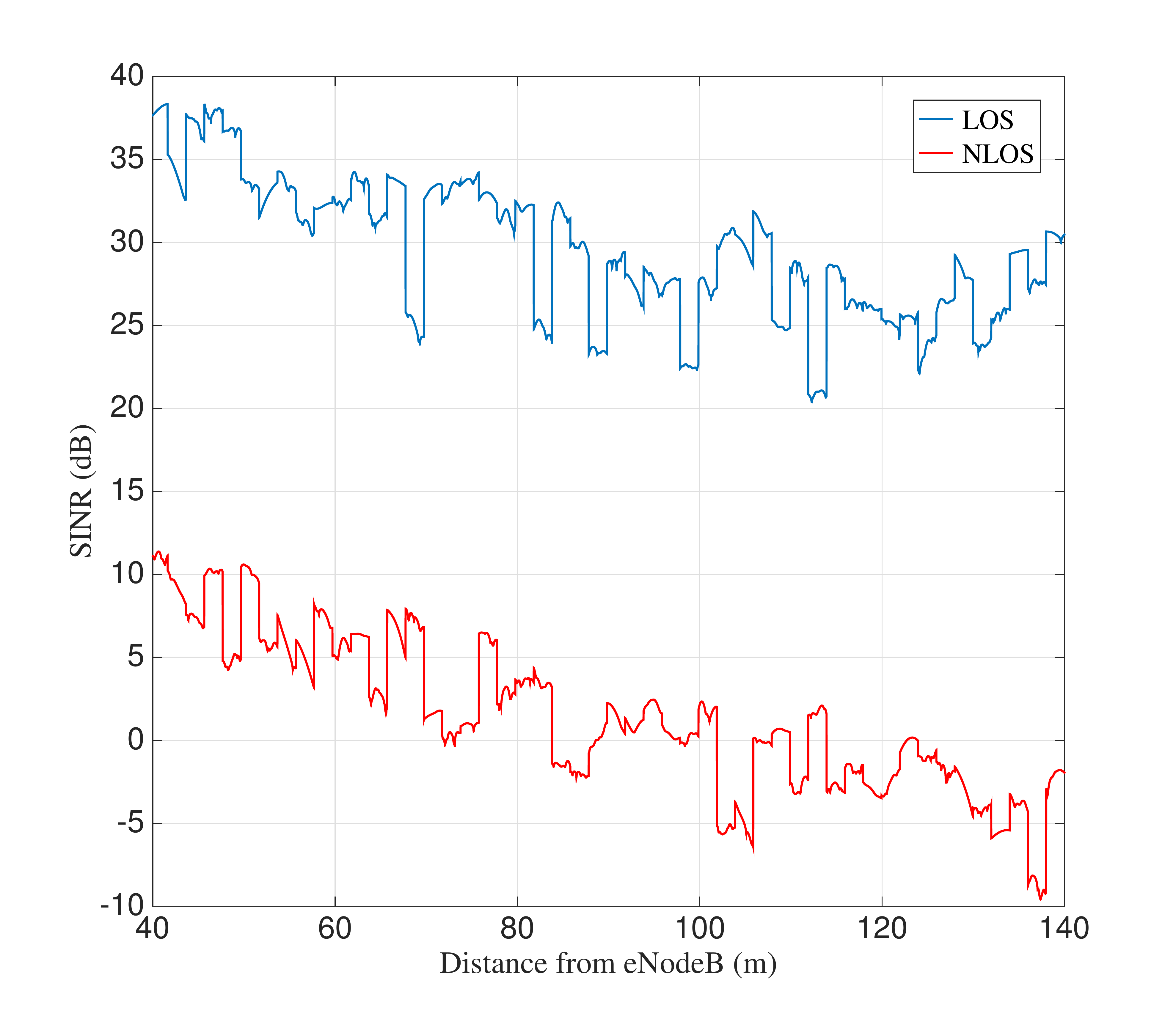}
                \caption{Average SINR estimated by the users.}
                \label{AvgSINR}
        \end{subfigure}
        \begin{subfigure}[b]{0.4\textwidth}
                \includegraphics[width=\textwidth]{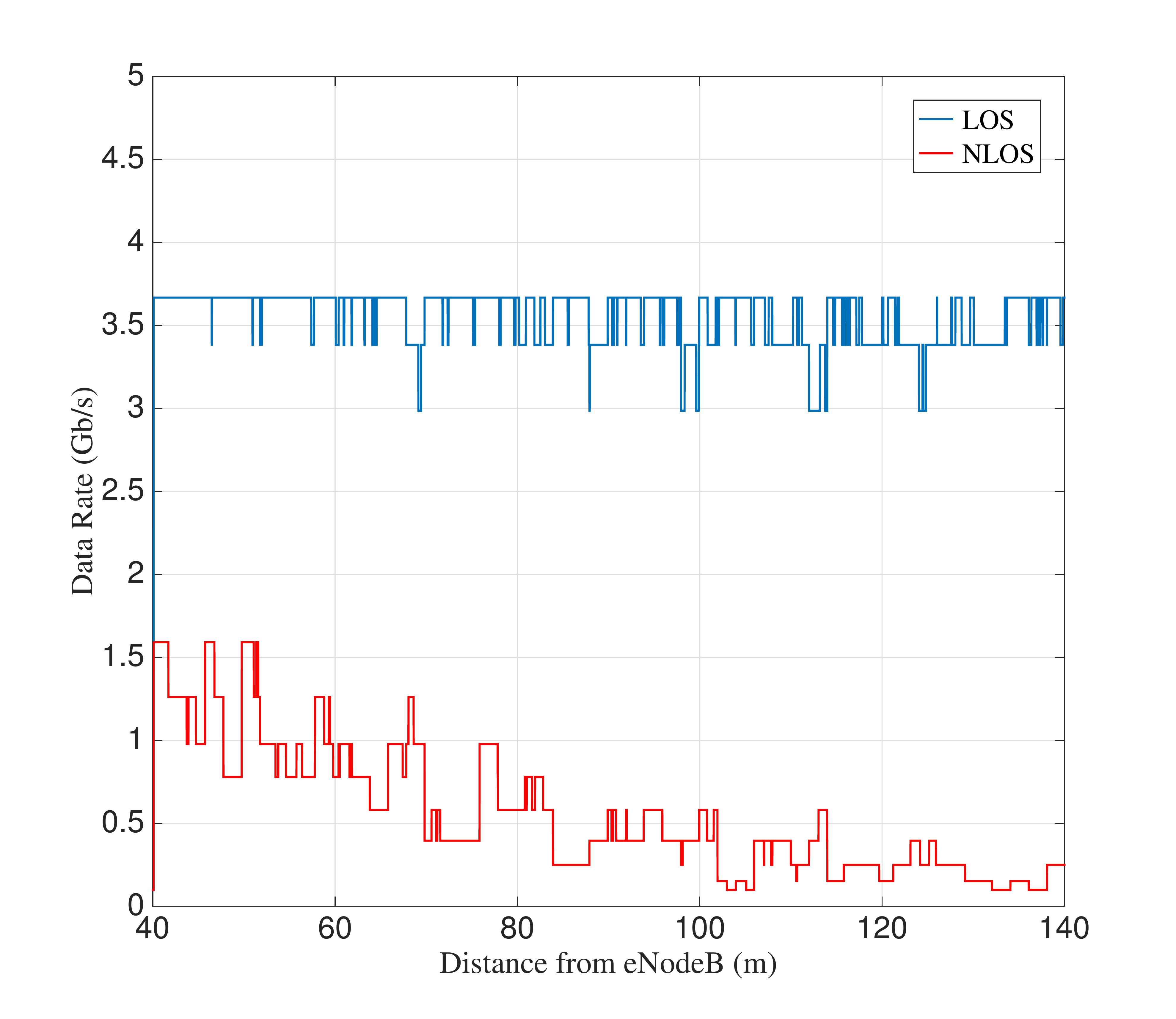}
                \caption{Data rate for users with LoS and NLoS links.}
                \label{DataRate}
        \end{subfigure}
       \vspace{2mm}
        \caption{\bf Simulation results}\label{SimRes}
\end{figure*}

\section{Simulation results}
\label{simuRes}
We validate our framework through a simple simulation scenario, where we consider one eNodeB and one UE. The user is at an initial distance of $40$m from the base station and moves towards the cell-edge with a constant velocity of $20$ m/s. The simulation configuration settings correspond to the default values shown in Table \ref{table1}. On top of that, the eNB has a transmit power of $30$ dBm and a receiver noise figure of $5$dB. Simulations are executed twice to capture the difference between the line of sight (LoS) and the non line of sight (NLoS) scenario.

The simulation results are reported in {\bf Fig. \ref{SimRes} }. On the one hand, {\bf Figs. \ref{SinrLos}} and {\bf \ref{SinrNlos}} show the variation of the downlink SINR with both time and frequency for the LoS and NLoS case, respectively. On the other hand, in {\bf Figs. \ref{DataRate}} and {\bf  \ref{AvgSINR} } we can observe the decrease in downlink data rate and average downlink SINR with the increasing distance between the UE and the base station, for both LoS and NLoS links. 

\section{Conclusions \& Future work}
\label{concl}
In this paper, a novel module for the simulation of mmWave cellular systems has been presented. The module, which is publicly available at \cite{github-ns3-mmwave}, is highly customizable to facilitate researchers to use it flexibly and analyze different scenarios using varying configurations. A basic implementation of mmWave devices, MAC layer,  PHY layer and channel models are developed. As part of our future work, we aim at introducing several enhancements, which include the \emph{integration} of (i) higher layer modules for end-to-end performance analysis, (ii) a HARQ module, (iii) uplink power control, (iv) ns-3 based channel matrix generation and beamforming computation, (v) more sophisticated MAC algorithms, (vi) relay devices and the \emph{evaluation} of (i) multiple access strategies and (ii) Transmission Control Protocol (TCP) performance over mmWave cellular networks.
\FloatBarrier

\balancecolumns

\begin{thebibliography} {20}

\bibitem {sundeep}
S. Rangan, T. S. Rappaport, and E. Erkip,
Millimeter-wave cellular wireless networks: potentials and challenges,
\emph{Proc. IEEE}, 102(3):366-385, Mar. 2014.

\bibitem{mustafa}
M. R. Akdeniz, Y. Liu, M. K. Samimi, S. Sun, S. Rangan, T. S. Rappaport, and E. Erkip,
mmWave channel modeling and cellular capacity evaluation,
\emph{IEEE J. Sel. Areas Commun.,} 32(6): 1164-1179, Jun. 2014.

\bibitem{ns3}  
ns-3 Network Simulator;
  \url{https://www.nsnam.org/}
  
\bibitem {ghosh2014millimeter}
A. Ghosh, T.A. Thomas, M.C. Cudak, R. Ratasuk, P. Moorut, F.W. Vook, T.S. Rappaport, G.R. MacCartney, S. Sun, and S. Nie,
mmWave enhanced local area systems: A high data rate approach for future wireless networks,
\emph{IEEE J. Sel. Areas Commun.}, 1152 - 1163, June 2014.

\bibitem {radio_interface}
T. Levanen, J. Pirskanen, and M. Valkama,
Radio Interface Design for Ultra-Low Latency Millimeter-Wave Communications in 5G Era,
\emph{IEEE GLOBECOM}, Dec. 2014.

\bibitem {samsung}
F. Khan, and J. Pi, 
mmWave mobile broadband: unleashing the 3-300GHz spectrum,
presented at \emph{IEEE Wireless Commun. Netw. Conf.}, Mar. 2011.

\bibitem{measurements}
T. S. Rappaport, S. Sun, R. Mazius, and H. Zhao, Y. Azar, K. Wang, G. N. Wong, J. K. Schulz, M. K. Samimi, and F. Gutierrez, Jr.,
Millimeter Wave Mobile Communications for {5G} Cellular: It Will Work!,
\emph{IEEE Access},
(1): 335-349, 2013.

\bibitem {marco}
M. Mezzavilla, M. Miozzo, M. Rossi, N. Baldo and M. Zorzi,
A Lightweight and Accurate Link Abstraction Model
for the Simulation of LTE Networks in ns-3,
\emph{IEEE MSWiM}, Oct. 2012.

\bibitem{FemtoForum}
FemtoForum, 
LTE MAC Scheduler Interface Specification v1.11,
Oct. 2010.

\bibitem{ns3LTE}
G. Piro, N. Baldo, and M. Miozzo,
An LTE module for ns-3 network simulator,
\emph{Proc. of Int. ICST Conf. on Simulation Tools and Techniques}, Mar. 2011.

\bibitem{github-ns3-mmwave}  
mmWave module for ns-3;
  \url{https://github.com/mmezzavilla/ns3-mmwave}
  
\end{thebibliography}
\end{document}